\RequirePackage{xspace}

\def\Dzero  {D\O\xspace}
\newcommand{\met}{\mbox{\ensuremath{\slash\kern-.7emE_{T}}}\xspace}
\newcommand{\mht}{\mbox{\ensuremath{\slash\kern-.7emH_{T}}}\xspace}

\documentclass[slac_one]{revtex4}
\usepackage{graphicx}
\usepackage{fancyhdr}
\def \D0 {D\O }
\begin{document}

\title{Searches for squarks and gluinos at \D0 } 

%

\author{Krisztian Peters (for the \D0 \  Collaboration)}
\affiliation{School of Physics and Astronomy, The University of Manchester,\\
Oxford Road, Manchester M13 9PL, United Kingdom}

\begin{abstract}
Recent results obtained by the \D0 \ Collaboration on searches for
squarks and gluinos in $p\bar p$ collisions at $\sqrt s =1.96$ TeV at
the Fermilab Tevatron collider are discussed. For the inclusive
searches, events with multiple jets of hadrons and large missing
transverse energy in the final state are studied within the framework
of minimal supergravity and assuming R-parity conservation. Searches
for the scalar top quark are studied in two final state topologies:
with events of two charm jets and large missing transverse energy or
leptons, jets and missing transverse energy. The data, corresponding
to integrated luminosities of up to 2.1 fb$^{-1}$, show no excess of a
signal above the expected background in any of the decay channels
examined. Instead, upper limits at 95\% Confidence Level on the
squarks and gluino masses are derived.
\end{abstract}

\maketitle

\thispagestyle{fancy}


\section{INTRODUCTION} 

Supersymmetric models \cite{susy} are amongst the most promising
fundamental theories to extend the Standard Model (SM) and describe
physics at arbitrarily high energies. Each of the SM particles has a
partner differing by a half-unit of spin. If R-parity \cite{rparity}
is conserved, supersymmetric particles are produced in pairs, and
their decay leads to SM particles and to the lightest supersymmetric
particle (LSP), which is stable. Cosmological arguments suggest that
the LSP should be neutral and colorless. The lightest neutralino
$\tilde{\chi}^0_1$, which is a mixture of the superpartners of the
neutral gauge and Higgs bosons, fulfills these conditions. In the
following, it is assumed that R-parity is conserved. Since the LSP is
weakly interacting, it escapes detection and provides the missing
transverse energy (\met) signature. In $p\bar p$ collisions, squarks
($\tilde{q}$) and gluinos ($\tilde{g}$), the superpartners of quarks
and gluons, would be abundantly produced, if sufficiently light, by
the strong interaction, leading to final states containing jets and
\met. The limits discussed in this note have been obtained in the
model of minimal supergravity (mSUGRA) \cite{msugra}.

\section{INCLUSIVE SEARCH FOR SQUARKS AND GLUINOS}

The inclusive searches for squarks and gluinos have been performed in
three separate analyses \cite{patrice}. The analysis strategy was to
optimize for three benchmark regions of the mSUGRA parameter space. A
``dijet'' analysis was optimized at low $m_0$ for events containing a
pair of acoplanar jets, as expected from $p\bar
p\to\tilde{q}\bar{\tilde{q}}\to q\tilde{\chi}^0_1\bar
q\tilde{\chi}^0_1$ and $p\bar p\to\tilde{q}\tilde{q}\to
q\tilde{\chi}^0_1 q\tilde{\chi}^0_1$.  A “gluino” analysis was
optimized at high $m_0$ for events with at least four jets, as
expected from $p\bar p\to\tilde{g}\tilde{g}\to q\bar q\tilde{\chi}^0_1
q\bar q\tilde{\chi}^0_1$. Finally, a “3-jets” analysis was
optimized for events with at least three jets, as expected from $p\bar
p\to\tilde{q}\tilde{g}\to q\tilde{\chi}^0_1 q\bar
q\tilde{\chi}^0_1$. The benchmark for this analysis is the case where
$m_{\tilde{q}} = m_{\tilde{g}}$.

The SM processes leading to events with jets and real \met in the
final state (``SM backgrounds'') are the production of $W$ or $Z$ bosons
in association with jets ($W/Z$+jets), of pairs of vector bosons
($WW$, $WZ$, $ZZ$) or top quarks ($t\bar t$), and of single top
quarks. The neutrinos from the decays $Z\to \nu\bar\nu$ and
$W\to\ell\nu$, with the $W$ boson produced directly or coming from a
top quark decay, generate the \met signature. While the $Z(\to
\nu\bar\nu)+$jets events are an irreducible background, most of the
$W$ boson leptonic decays leading to an electron or a muon were
identified, and the corresponding events rejected. However, a charged
lepton from $W$ boson decay can escape detection or fail the
identification criteria. Such $W$+jets events therefore exhibit the
jets plus \met signature. Finally, multijet production also leads to a
final state with jets and \met when one or more jets are mismeasured
(``multijet background'').

\begin{figure*}[t]
\centering
\includegraphics[width=100mm]{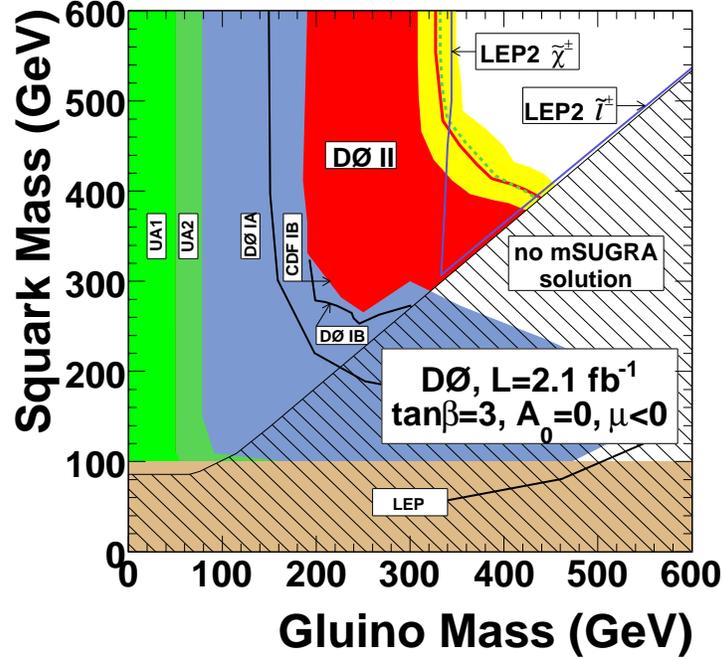}
\caption{In the gluino and squark mass plane, excluded regions
at the 95\% C.L. by direct searches in the mSUGRA framework with
$\tan\beta = 3$, $A_0 = 0$, $\mu < 0$. The thick (dotted) line is the
limit of the observed (expected) excluded region. Whereas, the yellow
band shows the effect of the PDF choice and of the variation of
renormalization and factorization scale by a factor of two.}
\label{sg1}
\end{figure*}

Each analysis required at least two jets and substantial \met ($\ge$
40~GeV). In the ``3-jets'' and ``gluino'' analyses, a third and fourth
jet were required, respectively. The transverse momenta of the jets up
to the third jet had to be greater than 35~GeV and greater than 20~GeV
for the fourth jet. The two leading jets (jets with largest transverse
momenta) were required: to be confirmed by tracks originating from the
primary vertex and to be in the central region of the calorimeter,
$|\eta_{det}| < 0.8$, where $|\eta_{det}|$ is the jet pseudorapidity
calculated from the detector center. The track confirmation
requirement removes events with spurious \met due to the choice of an
incorrect primary vertex. The acoplanarity, i.e. the azimuthal angle
between the two leading jets, was required to be smaller than
165$^0$. In all three analyses, a veto on isolated electrons or muons
with transverse momentum greater than 10~GeV was applied to reject
background events containing leptonic decays. The azimuthal angles
between the \met and the first and the second jet, ($\ge 90^0$ and
$\ge 50^0$ respectively), were used to remove events where the energy
of one jet was mismeasured, generating \met aligned with that jet. In
the ``dijet'' analysis, the minimum azimuthal angle $\Delta\phi (\met,
{\rm any~jet})$ between the \met and any jet with $p_T > 15$~GeV was
required to be greater than $40^0$ to further suppress the multijet
background.

The two final cuts on \met and on $H_T = \sum_{\rm jets} p_T$ , where
the sum is over all jets with $p_T > 15$~GeV and $|\eta_{det}| < 2.5$,
were optimized by minimizing the expected upper limit on the cross
section in the absence of signal. The minimum (\met, $H_T$) cuts were
(225, 325), (175, 375) and (100, 400) GeV for the ``dijet'', ``3-jets'' and
``gluino'' analysis respectively. The multijet background contribution
was estimated by fitting the \met distribution below 60~GeV with an
exponential function, after subtraction of the SM background
processes, and subsequently extrapolating this function above the
chosen \met cut value. This contribution was found to be negligible
and was conservatively ignored in the limit setting.

The numbers of events observed in the data are in agreement with the
SM background expectation in the three analyses. The number of events
passing at least one of the three analyses is 31 while the SM
expectation is $32.6\pm1.7$~(stat.)$^{+9.0}_{-5.8}$~(syst.) 
events. Fig.~1 shows the excluded domain at 95\% C.L. in the
gluino-squark mass plane with an integrated luminosity of 2.1
fb$^{-1}$.  The observed (expected) limits on the squark and gluino
masses are 392 (391)~GeV and 327 (332)~GeV. Limits were also derived
for the particular case of $m_{\tilde{q}} = m_{\tilde{g}}$ where
squarks and gluino masses below 390~GeV are excluded.



\section{SEARCH FOR SUPERSYMMETRIC TOP QUARK IN JETS AND \met FINAL STATE}

The mass splitting between the supersymmetric partners of the top
quark ($\tilde{t}$) is proportional to the mass of the top quark. This
can make one of the stops the lightest of all squarks. In this search
\cite{mansora}, the stop quark is taken to be the next to the lightest
supersymmetric particle and is assumed to decay into a charm quark and
a $\tilde{\chi}^0_1$ via loop decay. The final state consists of two
acoplanar charm jets and missing transverse energy.  The backgrounds
to this channel are similar as to the generic squarks and gluinos
search and the multijet background is estimated in the same way as
well, as described in the previous section.

\begin{figure*}[t]
\centering
\includegraphics[width=100mm]{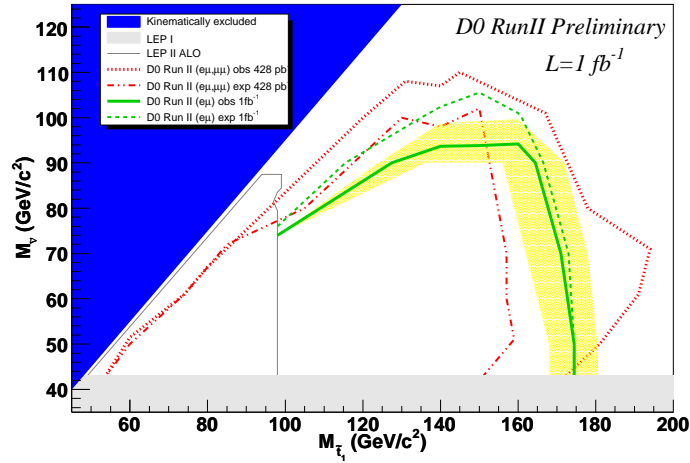}
\caption{Region in the $\tilde{t}$--$\tilde{\chi}^0_1$
mass plane excluded at the 95\% C.L. by the present search. The
observed (expected) exclusion contour is shown as the green solid
(dashed) line. The yellow band represents the theoretical uncertainty
on the scalar top quark pair production cross section due to PDF and
renormalization and factorization scale choice.} \label{cc}
\end{figure*}

The search strategy for $\tilde{t}$ involves three steps which include
the application of the selection criteria on kinematical variables,
heavy flavor tagging and optimization of the final selection depending
on $\tilde{t}$ and $\tilde{\chi}^0_1$ masses. The selection cuts
require exactly two jets, within $|\eta_{det}|$ of 1.5 and confirmed
by tracks originating from the primary vertex. The leading and
trailing jets have to have a transverse momentum more than 40 and
20~GeV respectively and separated by an azimuthal angle of less than
165$^0$. Events with isolated electrons, muons and tracks are removed
to reduce the backgrounds with $W$ decays. Several topological cuts on
the combination of \met and $H_T$ are further applied to reduce the
multijet contribution with mismeasured \met. To enhance the rate of
events with charm jets in the final state at least one of the jets is
required to be heavy flavor tagged with \Dzero's neural network (NN)
based tagging tool. The requirement on the NN output is kept loose to
get a high efficiency for detection of charm jets ($\approx$ 30\%)
with a $\approx$ 6\% probability for a light parton jet to be
mistakenly tagged.

At the final stage of the analysis, additional selection criteria on
three kinematic variables: \met , $H_T$ and $S =
\Delta\phi_{max}+\Delta\phi_{min}$, where $\Delta\phi_{max}$
($\Delta\phi_{min}$) is the largest (smallest) azimuthal separation
between a jet and \met, are optimized by maximizing the expected lower
limit on the neutralino mass for a given $m_{\tilde{t}}$. In all cases
a requirement of $\met > 70$~GeV is imposed. No contamination remains
from multijet background at this point in the analysis, which is
therefore neglected while setting the limit. Using the assumption that
$\tilde{t}$ decays into a charm quark and a neutralino with 100\%
branching fraction and the nominal $\tilde{t}$ pair production cross
section, the largest $m_{\tilde{t}}$ excluded by this analysis is
155~GeV, for a neutralino mass of 70~GeV at the 95\% C.L. The excluded
region in the $\tilde{t}$--$\tilde{\chi}^0_1$ mass plane with 1
fb$^{-1}$ of data is shown in Fig.~2.

\section{SEARCH FOR SUPERSYMMETRIC TOP QUARK IN
LEPTONS, JETS AND \met FINAL STATE}

\begin{figure*}[t]
\centering
\includegraphics[width=100mm]{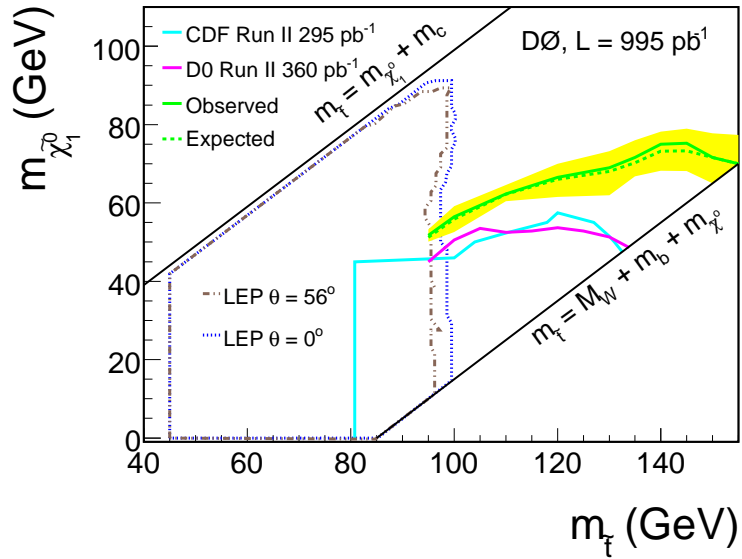}
\caption{Exclusion limits in the stop mass versus sneutrino mass 
plane at the 95\% C.L. The dashed green line is the expected limit
and the full green line is the observed limit. The yellow band around
the observed limit accounts for the effect of the stop cross section
uncertainties.} \label{emu}
\end{figure*}

If the $\tilde{t}\to c\tilde{\chi}^0_1$ and $\tilde{t}\to
b\tilde{\chi}^\pm_1$ modes are kinematically forbidden, then stop
quarks will decay into a lepton, a bottom quark and a sneutrino via
$\tilde{t}\to b\ell\tilde{\nu}$. The event signature for stop pair
production will be two b jets, a pair of isolated leptons and missing
transverse energy. A search is performed with 1.1 fb$^{-1}$ of \Dzero
data in the most sensitive channel where one lepton is a muon and the
other one an electron \cite{rest}, which has the largest branching
ratio and the smallest backgrounds among the three leptonic decay
modes.

The SM backgounds to this channel come from processes with isolated
electron and muon pairs: $Z\to\tau\tau$, $WW$, $WZ$, $ZZ$, and $t\bar
t$. The second background source is due to either mis-identified
electrons, muons or jets, mismeasured \met, or electrons or muons from
multijet processes which pass lepton isolation requirements. This
multijet background is estimated from data.

The selection cuts require events with an electron and a muon. Jets,
electron and muon have to be isolated from each other. The analysis is
using further a set of cuts on the azimuthal separation between the
electron, muon and \met to reduce the multijet background. The \met is
required to be larger than 30~GeV.

The exclusion limits have been derived from the shapes of the $H_T$
and $S_T$ distributions, where $S_T$ is defined as the scalar sum of
the electron, muon and missing transverse energy. The exclusion domain
is shown in Fig.~3 in the ($m_{\tilde{t}},m_{\tilde{\nu}}$) plane. For
large mass differences, stop masses lower than 175 GeV are excluded.

%

\end{document}